# Coalitional Game Based Carpooling Algorithms for Quality of Experience

(Technical Report)

Jiale Huang, Jigang Wu, Long Chen
School of Computer Science and Technology
Guangdong University of Technology, Guangzhou, China
Email: {fatjlhuang, asjgwucng}@outlook.com, lonchen@mail.ustc.edu.cn

Abstract—Carpooling service is an effective solution to balance the limited number of taxicabs and the soaring demands from users. Thus, how to motivate more passengers to participate in carpooling is essential, especially in extreme weather or in rush hours. Most of existing works focus on improving the availability, convenience and security of carpooling service, while ignoring to guarantee the quality of experience (QoE) of passengers. In this work, we focus on how to fulfill the expected sojourn time of passengers in carpooling service using coalition game. We formulate the QoE-guarantee problem as a benefit allocation problem. To solve the problem, we quantify the impatience of passengers due to detouring time delay, depending on their own expected sojourn time and expected compensation per unit time of delay. The algorithm named PCA is proposed to minimize the impatience of all passengers, under which we calculate the compensation for passengers based on Shapley value. We prove that PCA can guarantee the fairness of passengers. Extensive simulation results demonstrate that PCA can minimize the impatience of passengers. Moreover, compared with the existing algorithm DST, PCA can reduce the payment of each passenger by 14.4% averagely with only 13.3% loss of driver's revenue. However, the least expected revenue of the driver can still be fulfilled, which produces a win-win solution for both passengers and drivers in carpooling.

## I. Introduction

Urban traffic congestion can cause numerous detrimental effects, such as time loss, air pollution and excessive fuel consumption [1], [2]. Carpooling service is a beneficial solution to solve the problem. By allowing multiple passengers to share a vehicle, carpooling can reduce the total mileage to deliver passengers, therefore alleviating traffic congestion and reducing excessive emissions effectively [3]–[5]. On one hand, the ride demands from users explosively increase in rush hours. On the other hand, a

This work has been accepted by The 24th International Conference on Parallel and Distributed Systems December 11 - 13, 2018, Singapore.

taxicab only carries one or two passengers and leaves other seats empty [6]. By integrating empty seat resources, carpooling service can make more passengers share a taxicab [7], thus maintaining the balance between the limited taxicab resources and the soaring ride demands.

Many works on carpooling service focus on designing recommendation systems [5], [7]–[9], solving vehicle routing problems [9]–[12] and inventing security protocols [13]–[16] to improve availability, convenience and security of carpooling. However, only a few works have focused on designing incentive mechanisms to motivate both passengers and drivers to participate in carpooling. Zhang et al. [7] presented a reciprocal price mechanism to provide economic incentives for carpooling. Zhang et al. [9] proposed a win-win carpool fare model to encourage both drivers and passengers to participate in carpooling. Li et al. [17] proposed a dynamic pricing method and divide the payoffs according to the contribution of each individual. However, all the above studies cannot guarantee the quality of experience (QoE) of passengers. A simple example in the following paragraph is used to better illustrate this problem.

As shown in Fig. 1, passenger $p_1$ and passenger $p_2$ request for carpooling service to improve the probability for successfully taking a taxicab in rush hours. The start point of $p_1$ is close to the start point of $p_2$, and the destination of $p_1$ is close to the destination of $p_2$. Before carpooling service starts, $p_1$ sends route 1 to server shown in Fig. 2 while $p_2$ sends route 2 to the server. Then the server calculates the best route for carpooling, which is route 3. Driver will pick up $p_2$ before $p_1$ and put down $p_1$ before $p_2$ based on the routes on the map. When carpooling service completes, each passenger will arrive at destination. However, the actual travel time of passengers are prolonged due to detours, which will make them be impatient. Meanwhile, each passenger will still pay the fare based on his/her actual travel time and travel distance. Therefore, the QoE of them will suffer. With lower QoE, passengers will be unwilling to participate in carpooling. In this work, we propose a compensation scheme to ensure the QoE of passengers, i.e., to mitigate the impatience of them. Meanwhile, the proposed scheme can guarantee the interest of both passengers and drivers, so as to attract them to participate in carpooling. To effectively solve the QoE guarantee problem, we have to solve the following challenges: (I) Passengers have their own interests to decide whether to participate in carpooling or not. Therefore, how to make passengers have no incentive to reject to participate in carpooling is a challenge. (II) In carpooling service, the routes of all passengers are not exactly the same, which can make them suffer time delay, thus making them be impatient. We decide to compensate passengers based on the impatience of them suffered in carpooling.

Therefore, how to quantify the impatience of passengers is important. (III) Passengers should be treated fairly, which means that the compensation algorithm should ensure the fairness of passengers.

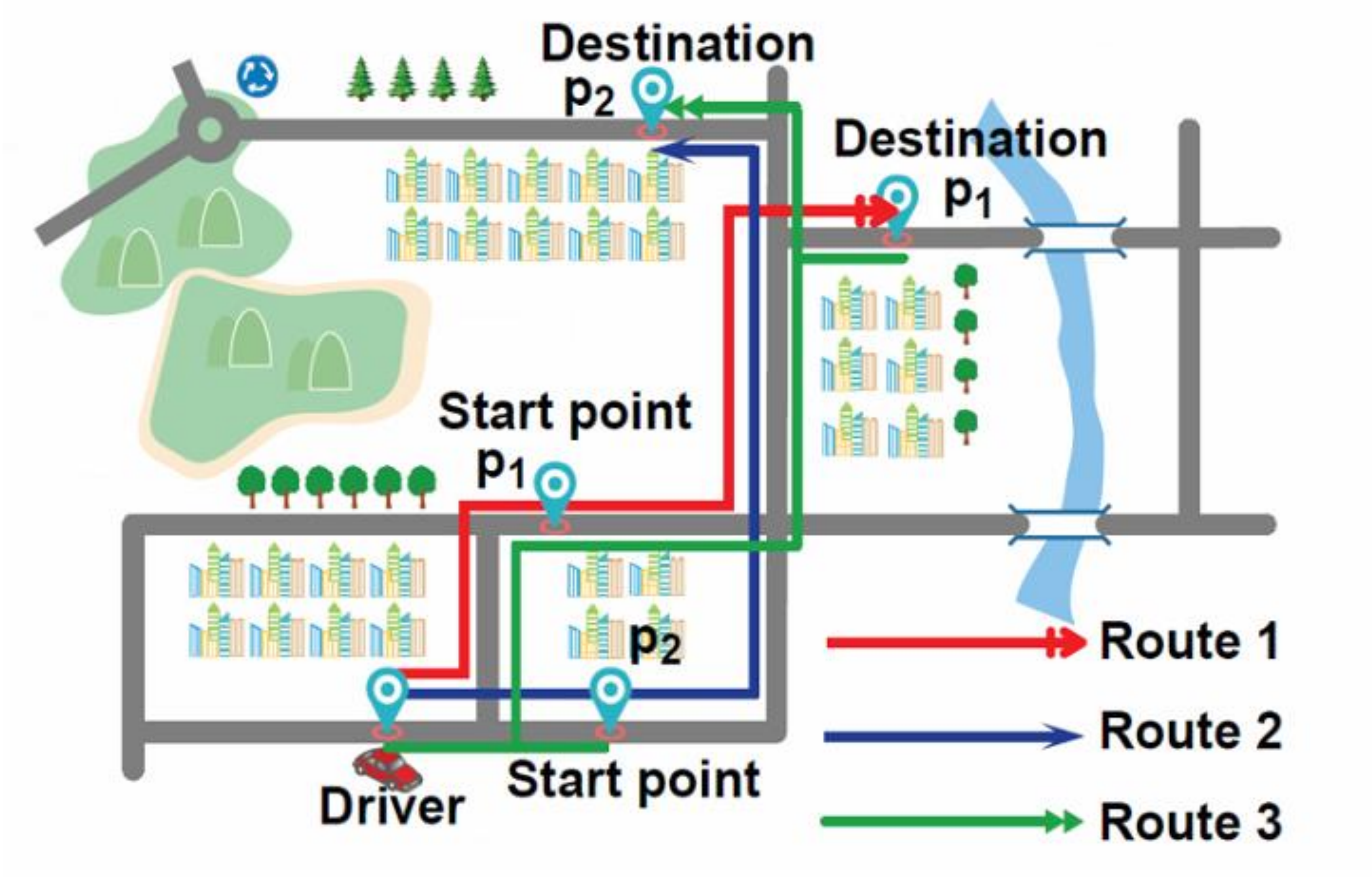

Fig. 1 Motivation Example

## II. Problem Formulation

TABLE I: Table of key notations

| Notation | Description |
|---|---|
| $d$ | Driver |
| $P$ | The set of passengers |
| $p_i$ | The i-th passenger |
| $T_i$ | The travel of the i-th passenger |
| $l_i$ | The travel distance of the i-th passenger |
| $t_i$ | The expected travel time of the i-th passenger |
| $pr_l$ | Price per kilometer |
| $pr_t$ | Price per minute |
| $x_d$ | The actual revenue of driver |
| $x_i$ | The compensation for the i-th passenger |
| $\theta_i$ | Expected sojourn time of the i-th passenger |
| $\omega_i$ | Expected compensation per unit time of delay of the i-th passenger |
| $\Sigma$ | The collection of all possible sequences in which driver serves passengers |

For the benefit of simplicity, the fare for a travel T with distance l and time t is calculated by [16]:

$$F(T) = pr_l * l + pr_t * t.$$

In rush hours, the limited number of taxicabs is difficult to meet passengers' demands. In this setting, incentive to motivate drivers to pick up more passengers is needed. Therefore, we define a price fluctuation coefficient ρ to give drivers more than general fare to motivate them. And the fare for a travel with price fluctuation coefficient is formulated as [16]:

$$G(T) = \rho * F(T).$$

Moreover, passengers have their own highest price αF(T) that they are willing to pay for their own travel. In fact, the fare for a travel is often no more than the highest price, and the difference between them is called passenger surplus, given by [16]:

$$U(p) = \alpha * F(T) - \rho * F(T).$$

Similarly, drivers have the least expected revenue βF(T) that they are willing to provide passengers with a taxicab service. The difference between the actual revenue of driver $x_d$ and the least expected revenue of driver is defined as driver surplus, given by:

$$U(d) = x_d - \beta * F(T).$$

In carpooling service, the routes of all passengers are not exactly the same so that the actual travel time of each passenger may be prolonged. Time delay may cause a decrease in the QoE of passengers, i.e., make them be impatient. Meanwhile, passengers have their own expected sojourn time and expected compensation per unit time of delay. Assume that the routes of all passengers are exactly different. When driver serve passengers following different sequence, passengers may suffer different degrees of impatience.

Given a sequence σ, we quantify the impatience of the i-th passenger suffered in carpooling as:

$$I_i(\sigma) = \theta_i * \omega_i + \omega_i * \sum_{j \in Pr_i(\sigma)} \theta_j.$$

And given a sequence σ, the impatience of all passengers suffered in carpooling is:

$$I(P, \sigma) = \sum_{i \in P} (\theta_i * \omega_i) + \sum_{i \in P} \left[ \omega_i * \sum_{j \in Pr_i(\sigma)} \theta_j \right].$$

For drivers, we should make them be willing to serve passengers. Therefore, we define an incentive coefficient ε which is bigger than β but no more than ρ, and use it to give driver a fare higher than his/her least expected revenue. And the actual revenue of the driver is calculated by:

$$x_d = \sum_{i \in P} \varepsilon * F(T_i).$$

The total compensation is calculated by:

$$G_{total} = \sum_{i \in P} G(T_i).$$

Therefore, we have:

$$x_d + \sum_{i \in P} x_i = \sum_{i \in P} G(T_i).$$

And the QoE-guarantee problem is formulated as:

$$
\begin{aligned}
OBJ: \quad & \max_{S \subseteq P} \left\{ \min_{i \in S} \frac{x_i}{I(S, \sigma^*)} \right\} \\
s.t. \quad & \sum_{i \in S} x_i = G_{total} - x_d, & (C1) \\
& x_d > 0, & (C2) \\
& x_d - \sum_{i \in S} \beta * F(T_i) \geq 0, & (C3) \\
& \alpha \geq \rho \geq \beta > 0, & (C4) \\
& x_i > 0, \; i \in S, & (C5) \\
& I(S, \sigma^*) \leq I(S, \sigma), \; \sigma \in \Sigma. & (C6)
\end{aligned}
$$

Where the objective function guarantees the QoE of passengers by compensating them with more fare. Constraint C1 illustrates the source of compensation for passengers, while constraints C2 and C3 indicate that the interest of drivers can be guaranteed. Constraints C4 and C5 ensure that carpooling is beneficial to passengers, while constraint C6 guarantees the impatience of passengers suffered in carpooling is minimal.

## III. Simulation Results

Using coalitional game, we have modeled carpooling service in rush hours as a coalitional game. By analyzing the utility of coalition formed by passengers and driver, we have proved that the coalition in carpooling service is stable. Depending on passengers' expected sojourn time and expected compensation per unit time of delay, we have quantified their impatience due to detouring time delay. We have proposed an effective compensation algorithm named PCA to minimize the impatience of all passengers suffered in carpooling, under which we compensate passengers based on Shapley value. We have proved that PCA can guarantee the fairness of passengers. Extensive simulation results indicate that, PCA can minimize the impatience of all passengers suffered in carpooling effectively. Moreover, PCA can reduce the payment of each passenger by 14.4% averagely with only 13.3% loss of driver's revenue, compared with DST [16]. However, PCA can give driver a fare higher than his/her least expected revenue, which means that PCA is a win-win solution for carpooling to guarantee both the interests of drivers and the QoE of passengers. In the future, we will try to extend our model to suit for the general case when there are multiple taxicabs.

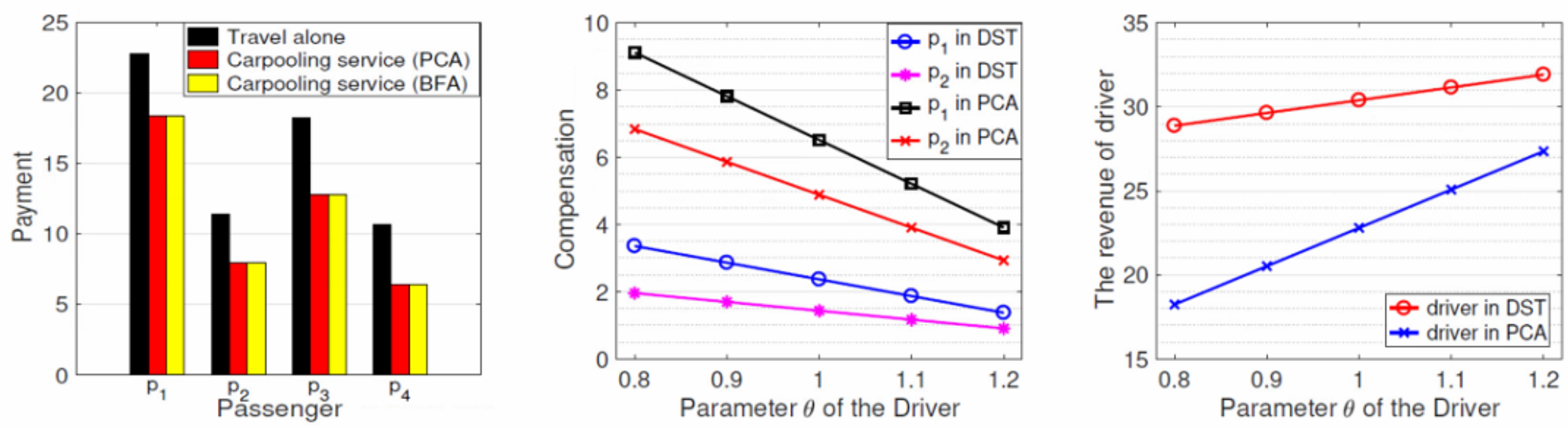

(a) Payments for carpooling and travel alone (b) Passengers (c) Driver

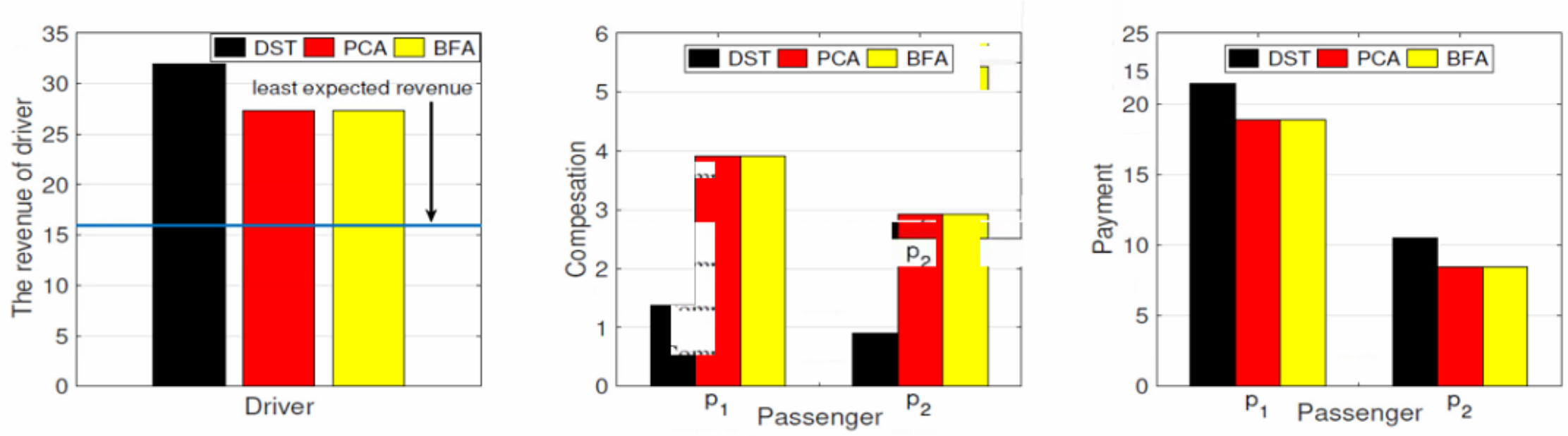

(a) The revenue of driver using DST and PCA (b) Compensation for each passenger (c) Payment of each passenger